\documentclass[12pt]{article}

\usepackage{graphicx}

\begin{document}
\baselineskip=16pt

\title{\addvspace{-10mm} {\normalsize \hfill CFNUL/99-07 }\\
{\normalsize \hfill hep-ph/9908206} \\
\addvspace{15mm} Plasma wave instabilities 
induced by neutrinos}
\author{Lu\'{\i}s Bento \\
{\normalsize \emph{Centro de F\'{\i}sica Nuclear da Universidade de Lisboa,} 
} \\
{\normalsize \emph{Av. Prof. Gama Pinto 2, 1649-003 Lisboa, Portugal}}}
\date{July 99}
\maketitle

\begin{abstract}
Quantum field theory is applied to study the interaction of an electron
plasma with an intense neutrino flux. A connection is established between
the field theory results and classical kinetic theory. The dispersion
relation and damping rate of the plasma longitudinal waves are derived in
the presence of neutrinos. It is shown that Supernova neutrinos are never
collimated enough to cause non-linear effects associated with a neutrino
resonance. They only induce neutrino Landau damping, linearly proportional
to the neutrino flux and $G_{\mathrm{F}}^{2}$. \newline
\strut \newline
\end{abstract}


\newpage

\section{Introduction}

Bingham \emph{et~al.}~\cite{bing94} have studied the interaction of a
neutrino beam with an electron plasma and concluded that the neutrino fluxes
produced in Supernovae are intense enough to cause plasma instabilities with
large growth rates. If true this would provide a physical mechanism of
energy transfer from the neutrinos to the plasma that might explain the
Supernova explosions. The interaction between neutrinos and plasma was
described with a ponderomotive force acting on the electrons and a neutrino
wave function obeying a naive Klein-Gordon equation modified with a matter
induced external potential. This non-standard treatment was not well
established from the Standard Model of electroweak interactions and
originated some controversy \cite{hard96,hard97}. More recently \cite{silv98}%
, classic kinetic theory was applied to study the neutrino-plasma system
where both neutrino and electron particles suffer each own ponderomotive
force. Again the lepton spin and chiral structure of the weak interactions
remain unnoticed. The dispersion relation derived for the plasma waves
differed from the one previously obtained in \cite{bing94}. Yet, the authors
reiterated the claim that the neutrinos produce non-linear effects for
certain resonant modes of plasma waves causing instabilities with large
growth rates proportional, not to $G_{\mathrm{F}}^{2}n_{\nu }$, but to a
smaller power of this quantity.

The aim of the present work presented here is to obtain a formulation of the
problem based on field theory and Standard Model of electroweak
interactions, an effort initiated in \cite{bent98}. We confine to isotropic
plasmas and longitudinal photon excitations (also called plasmons or
Langmuir waves). The \v{C}erenkov emission of longitudinal photons by \emph{%
massless} Standard Model neutrinos in an isotropic plasma has been studied
along the years \cite{orae86,kirz90,sawy92,oliv96,sahu97}. This is a single
neutrino decay but not a collective neutrino process. Hardy and Melrose \cite
{hard97} gave one step more by extending the work to spontaneous and \emph{%
stimulated\ emission and absorption} of plasmons to study the so-called
kinetic instabilities (see also \cite{tsyt64}). They obtained an expression
for the decay\ (growth) rate of the plasma waves induced by a neutrino flux.
However, because the derivation was based on single neutrino processes the
result is necessarily proportional to $G_{\mathrm{F}}^{2}$ excluding \emph{%
apriori} any possible non-linear effects. In the present paper we derive the
neutrino contribution to the photon self-energy and obtain a modified
dispersion relation for the longitudinal waves. This allows one to study not
only kinetic instabilities but also the possible existence of hydrodynamic
instabilities and non-linear phenomena.

Next section we calculate the neutrino contribution to the electromagnetic
polarization tensor and the dispersion relation of longitudinal photons. In
section 3 we establish a relationship with kinetic theory and in section 4
the dispersion relation and nature of neutrino induced instabilities are
analyzed in detail. In the last section we summarize the main results.

\section{Electromagnetic polarization and longitudinal waves}

\begin{figure}[b]
\centering
\vspace{10pt}
\includegraphics*[width=36mm]{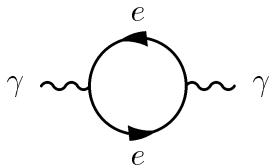}
\caption{Electron plasma contribution to the photon self-energy, 
$i\pi _{\mathrm{EM}}^{\mu \nu }$.}
\label{fig1}
\end{figure}

In a medium the Maxwell equations are modified by polarization effects. The
electromagnetic waves obey the following equation: 
\begin{equation}
\left( -k^{2}g^{\mu \nu }+k^{\mu }k^{\nu }+\pi ^{\mu \nu }\right)
\varepsilon _{\nu }=0\;,  \label{wave}
\end{equation}
where $\varepsilon $ is the wave polarization vector, $k$ the linear
momentum and $\pi ^{\mu \nu }$ is the polarization tensor. The purely
electromagnetic contribution, $\pi _{\mathrm{EM}}^{\mu \nu }$, is well known
for an electron plasma in first order of approximation and can be identified
with the Feynman diagram of Fig.~\ref{fig1}. At finite temperature the one-particle
propagators possess additional terms related to the one-particle
distribution functions of the background matter. In the real-time formalism 
\cite{dola74,weld82,land87}, adopted here, the electron propagator in a homogeneous
and unpolarized plasma is given by 
\begin{equation}
(p\hspace{-0.19cm}/+m)\left[ \frac{i}{p^{2}-m^{2}}-2\pi \,\delta
(p^{2}-m^{2})\left( \theta (p^{0})f_{e}(p)\,+\theta (-p^{0})f_{\bar{e}%
}(-p)\right) \right] \;,  \label{epro}
\end{equation}
where $f_{e},$ $f_{\bar{e}}$ are the electron and positron distribution
functions respectively.

Gauge invariance implies that $\varepsilon $ is defined up to a vector
proportional to $k$ and $k_{\mu }\pi ^{\mu \nu }=0=\pi ^{\mu \nu }k_{\nu }$.
On the other hand, for an isotropic, homogeneous and unpolarized plasma the
tensor $\pi _{\mathrm{EM}}^{\mu \nu }$ is symmetric and can be written in
terms of the metric tensor, momentum $k$ and vector $u$ defining the time
direction in the plasma rest frame \cite{weld82}. As a result, 
\begin{equation}
\pi _{\mathrm{EM}}^{i0}=\frac{\omega k^{i}}{\vec{k}^{2}}\pi _{\mathrm{EM}%
}^{00}\;  \label{pii0}
\end{equation}
in the plasma frame ($\omega =k^{0}$) and the wave Eqs.~(\ref{wave}) admit a
purely electrostatic solution, $\varepsilon ^{\mu }=(1,\vec{0})=u^{\mu }$ in
the Coulomb gauge. The dispersion relation of these longitudinal waves is 
\begin{equation}
\vec{k}^{2}+\pi _{\mathrm{EM}}^{00}=0\;.
\end{equation}

\begin{figure}[t]
\centering
\includegraphics*[width=61mm]{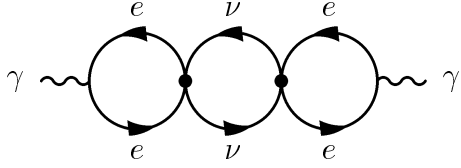}
\caption{Neutrino-plasma contribution to the photon self-energy, 
$i\pi _{\mathrm{W}}^{\mu \nu }$. }
\label{fig2}
\end{figure}

In the low energy limit of the Standard Model of electroweak interactions
the electron and neutrino interactions (both charged and neutral currents)
are described by the effective Lagrangian 
\begin{equation}
\mathcal{L}_{\mathrm{int}}=e\,A_{\mu }\,\bar{e}\gamma ^{\mu }e-\sqrt{2}\,G_{%
\mathrm{F}}\,(\bar{\nu}_{L}\gamma _{\mu }\nu _{L})\,\bar{e}\,\gamma ^{\mu
}(c_{V}^{\prime }\,-c_{A}^{\prime }\,\gamma _{5})e\,\ ,  \label{Leff}
\end{equation}
where $e$ denotes the positron charge ($\alpha =e^{2}/4\pi $), $G_{\mathrm{F}%
}$ the Fermi constant and 
\begin{equation}
c_{V}^{\prime }=c_{A}^{\prime }+2\sin ^{2}\theta _{\mathrm{W}}\;,
\end{equation}
with $c_{A}^{\prime }$ equal to $+1/2$ for $\nu _{e}$ and $-1/2$ for $\nu
_{\mu }$, $\nu _{\tau }$. The neutrino flux contributes to the
electromagnetic polarization through the diagram of Fig.~\ref{fig2}. The diagrams
with fermion self-energy corrections either in vacuum or in matter will be
neglected as well as the weak interactions between electrons or nucleons of
the medium. The expression of the $\nu _{L}$ propagators is similar to the
electron propagator of Eq.~(\ref{epro}): 
\begin{equation}
G_{\nu }(p)=\frac{1-\gamma _{5}}{2}\,\,p\hspace{-0.19cm}/\left[ \frac{i}{%
p^{2}}-2\pi \,\delta (p^{2})\left( \theta (p^{0})f_{\nu }(p)\,+\theta
(-p^{0})f_{\bar{\nu}}(-p)\right) \right] \;.  \label{npro}
\end{equation}
The only differences are that the $\nu $ masses are taken to be zero, there
is only one spin degree of freedom, $\nu _{L}$ or $\bar{\nu}_{R}$, and the
neutrino and anti-neutrino distribution functions, $f_{\nu }$ and $f_{\bar{%
\nu}}$, are not thermal, as they move along a privileged direction.

\begin{figure}[t]
\centering
\includegraphics*[width=32mm]{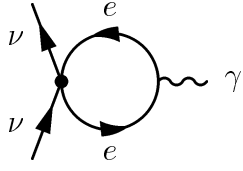}
\caption{Neutrino electromagnetic coupling induced by an electron plasma.}
\label{fig3}
\end{figure}

The neutrino loop gives a tensor, 
\begin{equation}
-i\,\pi _{\alpha \beta }^{N}(k)=-\int \frac{d^{4}p}{(2\pi )^{4}}\mathrm{Tr}%
\left\{ \gamma _{\alpha }G_{\nu }(p+k)\gamma _{\beta }G_{\nu }(p)\right\} \;,
\end{equation}
whose vacuum contribution is to be ignored. Only the contributions from the
electron and neutrino loops that are linear in the respective particle
densities will be retained. The electron loops are directly related to the
neutrino electromagnetic coupling. 
Writing the $\nu \nu \gamma $ vertex in Fig.~\ref{fig3} as 
\begin{equation}
-i\,\gamma _{\nu }\frac{1-\gamma _{5}}{2}\, \Gamma ^{\nu \mu }(k)\;,
\end{equation}
where $k$ is the incoming photon momentum, the diagram of Fig.~\ref{fig2} is given by 
\begin{equation}
i\,\pi _{\mathrm{W}}^{\mu \nu }=i\,\Gamma ^{\alpha \mu }(-k)\pi _{\alpha
\beta }^{N}(k)\Gamma ^{\beta \nu }(k)\;.  \label{piw}
\end{equation}
The $\nu \nu \gamma $ vertex separates in a pseudo-tensor proportional to $%
c_{A}^{\prime }$ and a tensor that is proportional to $\pi _{\mathrm{EM}%
}^{\mu \nu }$, as follows \cite{orae87,niev94}: 
\begin{equation}
\Gamma ^{\mu \nu }(k)=\frac{1}{e}\sqrt{2}G_{\mathrm{F}}\left( c_{V}^{\prime
}\pi _{\mathrm{EM}}^{\mu \nu }-c_{A}^{\prime }\pi _{5}\,\varepsilon ^{\mu
\nu \alpha \beta }k_{\alpha }u_{\beta }\right) \;.  \label{pi5}
\end{equation}
For future reference let us write the expressions of the susceptibility
tensors, 
\begin{equation}
\pi _{\mathrm{EM}}^{\mu \nu }=-2e^{2}\int \frac{d^{3}p}{(2\pi )^{3}}\frac{%
f_{e}+f_{\bar{e}}}{E_{e}}\frac{(k^{\mu }p^{\nu }+p^{\mu }k^{\nu }-k\!\cdot
\!p\,g^{\mu \nu })k\!\cdot \!p-k^{2}p^{\mu }p^{\nu }}{(k\!\cdot
\!p)^{2}-(k^{2}/2)^{2}}\;,  \label{piem}
\end{equation}
\begin{eqnarray}
\pi _{N}^{\mu \nu } &=&-\int \frac{d^{3}p}{(2\pi )^{3}}\frac{f_{\nu }+f_{%
\bar{\nu}}}{E_{\nu }}\frac{(k^{\mu }p^{\nu }+p^{\mu }k^{\nu }-k\!\cdot
\!p\,g^{\mu \nu })k\!\cdot \!p-k^{2}p^{\mu }p^{\nu }}{(k\!\cdot
\!p)^{2}-(k^{2}/2)^{2}}  \nonumber \\
&&-\frac{i}{2}\,\varepsilon ^{\mu \nu \alpha \beta }k_{\alpha }p_{\beta
}\int \frac{d^{3}p}{(2\pi )^{3}}\frac{f_{\nu }-f_{\bar{\nu}}}{E_{\nu }}\frac{%
k^{2}}{(k\!\cdot \!p)^{2}-(k^{2}/2)^{2}}\;.  \label{pin}
\end{eqnarray}

If the neutrino distributions were isotropic the polarization tensor $\,\pi
_{\mathrm{W}}$ would satisfy a relation like (\ref{pii0}) and the
longitudinal photons could still be described with a scalar potential \emph{%
i.e.}, $\varepsilon ^{\mu }=(1,\vec{0})$ in the plasma rest frame. That is
not strictly the case in the presence of a neutrino flux but since the
plasma is isotropic, as considered here, the vector components $\varepsilon
^{i}$ are expected to be suppressed with respect to $\varepsilon ^{0}$ by a
factor of $G_{\mathrm{F}}^{2}\,k^{4}$. We now prove that the components $%
\varepsilon ^{i}$ are in fact suppressed and their contribution to the
dispersion relation only comes at $G_{\mathrm{F}}^{4}$ order. First, note
that the assumed plasma isotropy implies not only the relation (\ref{pii0})
but also the structure \cite{weld82} 
\begin{equation}
\pi _{\mathrm{EM}}^{ij}=-\pi _{T}P_{ij}+\pi _{\mathrm{EM}}^{00}\frac{\omega
^{2}}{\vec{k}^{2}}\frac{k^{i}k^{j}}{\vec{k}^{2}}\;,
\end{equation}
where $P$ is the projector over the plane orthogonal to $\vec{k}$, 
\begin{equation}
P_{ij}=\delta _{ij}-\frac{k^{i}k^{j}}{\vec{k}^{2}}\;.
\end{equation}
Due to gauge invariance only three of the wave Eqs.~(\ref{wave}) are
linearly independent. Working in the Coulomb gauge, $\vec{k}\!\cdot \!\vec{%
\varepsilon}=0$, the application of the projector $P$ on the left of the
last three Eqs.~(\ref{wave}) gives 
\begin{equation}
\left( k^{2}-\pi _{T}\right) \varepsilon ^{i}=P_{ij}\,\pi _{\mathrm{W}%
}^{j\nu }\,\varepsilon _{\nu }\;.
\end{equation}
The factor $k^{2}-\pi _{T}$ vanishes for transverse photons but not for
longitudinal waves: in a non-relativistic gas, $\pi _{T}$ is equal to $%
\omega _{p}^{2}$, the square plasma frequency, and $k^{2}=\omega ^{2}-\vec{k}%
^{2}$. The above equation can be used to calculate $\varepsilon ^{i}$ in a
iterative way, as follows: 
\begin{equation}
\varepsilon ^{i}=P_{ij}\frac{\pi _{\mathrm{W}}^{j0}}{k^{2}-\pi _{T}}%
\varepsilon ^{0}+\cdots \;.
\end{equation}
It is clear that $\varepsilon ^{i}$ are suppressed by $G_{\mathrm{F}}^{2}$
with respect to $\varepsilon ^{0}$. When substituted in the first of the
wave Eqs.~(\ref{wave}), 
\begin{equation}
\left( \vec{k}^{2}+\pi _{\mathrm{EM}}^{00}+\pi _{\mathrm{W}}^{00}\right)
\varepsilon ^{0}=\pi _{\mathrm{W}}^{0i}\,\varepsilon ^{i}\;,
\end{equation}
it becomes evident that the vector components $\varepsilon ^{i}$ only
reflect in the dispersion relation at $G_{\mathrm{F}}^{4}$ order and can
thus be safely neglected. The dispersion relation of longitudinal waves is
so 
\begin{equation}
\vec{k}^{2}+\pi _{\mathrm{EM}}^{00}+\pi _{\mathrm{W}}^{00}=0  \label{drw}
\end{equation}
at $G_{\mathrm{F}}^{2}$ order. It remains to evaluate $\pi _{\mathrm{EM}%
}^{00}$ and $\pi _{\mathrm{W}}^{00}$.

The electroweak interactions conserve the electric charge and lepton numbers 
$L_{e}$, $L_{\mu }$, $L_{\tau }$. This implies that at energies considered
here, far below the muon mass, the electron, $\nu _{e}$, $\nu _{\mu }$ and $%
\nu _{\tau }$ numbers are separately conserved which materializes in the
conservation laws 
\begin{equation}
k_{\mu }\pi _{N}^{\mu \nu }=k_{\mu }\Gamma ^{\mu \nu }=0\;=\pi _{N}^{\mu \nu
}k_{\nu }=\Gamma ^{\mu \nu }k_{\nu }\;.
\end{equation}
Using repeatedly these relations together with Eq.~(\ref{pii0}) one obtains
from Eqs.~(\ref{piw}-\ref{pi5}) 
\begin{equation}
\pi _{\mathrm{W}}^{00}=\frac{2G_{\mathrm{F}}^{2}c_{V}^{\prime 2}}{e^{2}}%
\left( \pi _{\mathrm{EM}}^{00}\right) ^{2}\left( 1-\frac{\omega ^{2}}{\vec{k}%
^{2}}\right) ^{2}\pi _{N}^{00}\;.  \label{piw00}
\end{equation}
The outcome is that the $c_{A}^{\prime }$ coupling does not contribute to
the dispersion relation of longitudinal waves in an isotropic electron
plasma. On the other hand, $c_{V}^{\prime \,2}$ is much smaller for $\nu
_{\mu }$, $\nu _{\tau }$ ($c_{V}^{\prime }\approx -0.04$) than for $\nu _{e}$
($c_{V}^{\prime }\approx 0.96$) so only $\nu _{e}$ and $\bar{\nu}_{e}$ will
be taken in consideration.

The expressions of the response functions are 
\begin{equation}
\pi _{\mathrm{EM}}^{00}=-2e^{2}\int \frac{d^{3}p_{e}}{(2\pi )^{3}}\left(
f_{e}+f_{\bar{e}}\right) E_{e}\frac{\vec{k}^{2}\!-(\vec{k}\!\cdot \!\vec{v}%
_{e})^{2}}{(k\!\cdot \!p_{e})^{2}-(k^{2}/2)^{2}}\;,  \label{piem00}
\end{equation}
\begin{equation}
\pi _{N}^{00}=-\int \frac{d^{3}p_{\nu }}{(2\pi )^{3}}\left( f_{\nu }+f_{\bar{%
\nu}}\right) E_{\nu }\frac{\vec{k}^{2}\!-(\vec{k}\!\cdot \!\vec{v}_{\nu
})^{2}}{(k\!\cdot \!p_{\nu })^{2}-(k^{2}/2)^{2}}\;,  \label{pin00}
\end{equation}
where $E$ and $\vec{v}$ denote in each case the energy and velocity of the
particle. The factor of $2$ in front of the electromagnetic susceptibility
stands for the number of electron spin states. The normalization of the
distribution functions is fixed by the propagators (\ref{epro}) and (\ref
{npro}) as follows: the number of particles ($e^{-},$ $e^{+}$, $\nu $ or $%
\bar{\nu}$) per unity of volume \emph{and} spin degree of freedom is 
\begin{equation}
n=\int \frac{d^{3}p}{(2\pi )^{3}}f\;.
\end{equation}

Next section we seek a connection between field theory and classical kinetic
theory in the limit of small frequencies and wavelengths. In section 4 we
study the dispersion relation and possible neutrino induced instabilities.

\section{Kinetic Theory}

The Feynman diagrams in Figs.~1 and 2 are a consequence of a certain field
dynamics that couples boson fields with fermion densities. The electroweak
interactions in particular only involve vector fields and currents. The ones
relevant here are the electromagnetic field and current, $A_{\mu }$ and $J_{%
\mathrm{EM}}^{\mu }$, plus the neutrino and weak electron currents defined
as 
\begin{eqnarray}
J_{N}^{\mu } &=&\bar{\nu}_{L}\gamma ^{\mu }\nu _{L}\;, \\
J_{\mathrm{W}e}^{\mu } &=&\sqrt{2}G_{\mathrm{F}}\,\bar{e}\,\gamma ^{\mu
}(c_{V}^{\prime }\,-c_{A}^{\prime }\,\gamma _{5})e\;.
\end{eqnarray}
Behind those Feynman diagrams there is a set of equations relating the
current fluctuations, $J^{\mu }(k)$ in momentum space, to each other: 
\begin{eqnarray}
J_{\mathrm{EM}}^{\mu } &=&-\pi _{\mathrm{EM}}^{\mu \nu }(k)\,A_{\nu
}+J_{N\alpha }\Gamma ^{\alpha \mu }(-k)\;,  \label{j1} \\
J_{N\alpha } &=&-\pi _{\alpha \beta }^{N}(k)\,J_{\mathrm{W}e}^{\beta }\;,
\label{j2} \\
J_{\mathrm{W}e}^{\beta } &=&\Gamma ^{\beta \nu }(k)\,A_{\nu }+\pi _{\mathrm{W%
}e}^{\beta \nu }(k)\,J_{N\nu }\;.  \label{j3}
\end{eqnarray}
Here, $\pi _{\mathrm{W}e}$ is an electron loop, suppressed by $G_{\mathrm{F}%
}^{2}$, that couples the weak electron and neutrino currents. One obtains
from this a relation between the electromagnetic current and $A_{\mu }$.
Using matrix notation and the definition $\Gamma ^{\prime \mu \alpha }(k)=$ $%
\Gamma ^{\alpha \mu }(-k)$ one gets 
\begin{eqnarray}
J_{\mathrm{EM}} &=&-(\pi _{\mathrm{EM}}+\Gamma ^{\prime }\pi _{N}(1+\pi _{%
\mathrm{W}e}\pi _{N})^{-1}\Gamma )\,A \\
&\simeq &-(\pi _{\mathrm{EM}}+\Gamma ^{\prime }\pi _{N}\Gamma )\,A\;,
\end{eqnarray}
where the corrections of $G_{\mathrm{F}}^{4}$ order associated with $\pi _{%
\mathrm{W}e}$ are neglected in the last equation. When this electromagnetic
current is put in the Maxwell equations, 
\begin{equation}
\left( -k^{2}g^{\mu \nu }+k^{\mu }k^{\nu }\right) A_{\nu }=J_{\mathrm{EM}%
}^{\mu }\;,
\end{equation}
they give rise to the wave equations (\ref{wave}) with a polarization tensor
given by the Feynman diagrams of Figs.~1 and 2.

We have seen in the previous section that for an isotropic plasma the $%
c_{A}^{\prime }$ coupling and electron axial-current do not play a
significant role. If that current is dropped out of Eqs.~(\ref{j1}-\ref{j3})
they reduce to 
\begin{eqnarray}
J_{\mathrm{EM}}^{\mu } &=&\frac{1}{e}\pi _{\mathrm{EM}}^{\mu \nu }\left(
-eA_{\nu }+\sqrt{2}G_{\mathrm{F}}c_{V}^{\prime }\,J_{N\nu }\right) \;,
\label{JEM} \\
J_{N\alpha } &=&\frac{1}{e}\,\pi _{\alpha \beta }^{N}\,\left( \sqrt{2}G_{%
\mathrm{F}}c_{V}^{\prime }\,J_{\mathrm{EM}}^{\beta }\right) \;.  \label{JN}
\end{eqnarray}
which now have a classical counterpart in the sense that the fermion
polarization does not appear explicitly. This is the kind of relations also
obtained in classical kinetic theory, a framework that has been used by some
authors \cite{silv98,lami99}. That fact motivated us to formulate our own
classic theory in order to better understand the differences between the
results based on quantum field theory and their works.

In kinetic theory a system is described by distribution functions on the
phase space of the particles, in particular, the single particle
distribution functions $f(t,\vec{x},\vec{p})$. In low dense plasmas the
collisions are less important than the collective interactions and the time
evolution of the distribution functions is given by the Vlasov equations 
\cite{lifs97}, 
\begin{equation}
\frac{\partial f}{\partial t}+\vec{v}\cdot \!\frac{\partial f}{\partial \vec{%
x}}+\vec{F}\!\cdot \!\frac{\partial f}{\partial \vec{p}}=0\ .  \label{Dfen}
\end{equation}
The functions velocity, $\vec{v}$,\ and force, $\vec{F}$, have to be
specified for each of the particles species.

If one ignores the electron polarization and $c_{A}^{\prime }$ coupling, the
effective Lagrangian of Eq.~(\ref{Leff}) reduces to 
\begin{equation}
\mathcal{L}_{\mathrm{int}}=e\,A_{\mu }J_{e}^{\mu }-\sqrt{2}G_{\mathrm{F}%
}c_{V}^{\prime }\,J_{N\mu }J_{e}^{\mu }\;.
\end{equation}
It clearly admits a classical limit where the neutrino vector current $%
J_{N}^{\mu }$ is equal to the difference between the $\nu $ and $\bar{\nu}$
current densities, $J_{N}^{\mu }=j_{\nu }^{\mu }-j_{\bar{\nu}}^{\mu }$, and $%
J_{e}^{\mu }=j_{e}^{\mu }-j_{\bar{e}}^{\mu }$ is the analogous current for
electrons and positrons. The next step is to write down the interaction
Lagrangian for a classical electron or neutrino particle respectively \cite
{bent98}, 
\begin{eqnarray}
L_{e} &=&\left( e\,A^{\mu }-\sqrt{2}G_{\mathrm{F}}c_{V}^{\prime
}\,J_{N}^{\mu }\right) \dot{x}_{\mu }\;,  \label{Le} \\
L_{\nu } &=&-\left( \sqrt{2}G_{\mathrm{F}}c_{V}^{\prime }\,J_{e}^{\mu
}\right) \dot{x}_{\mu }\;.  \label{Ln}
\end{eqnarray}
They are symmetric to the positron and anti-neutrino Lagrangians. Notice
that the vector current $J_{N}^{\mu }$ couples to an electron particle in
exactly the same way as the electromagnetic potential $A_{\mu }$ and in turn
the neutrinos interact with a vector potential as well, proportional to $%
J_{e}^{\mu }$. Therefore, the electroweak forces are a straightforward
generalization of the electromagnetic Lorentz force \cite{bent98}. The total
force acting on an electron is 
\begin{equation}
\vec{F}_{e}=-e(\vec{E}+\vec{v}_{e}\wedge \vec{B})+\sqrt{2}G_{\mathrm{F}%
}c_{V}^{\prime }\,(\vec{E}_{N}+\vec{v}_{e}\wedge \vec{B}_{N})\;\,  \label{Fe}
\end{equation}
with weak-electric and weak-magnetic fields given by 
\begin{equation}
\vec{E}_{N}=-\vec{\nabla}J_{N}^{0}-\frac{\partial \vec{J}_{N}}{\partial t}%
\;,\qquad \vec{B}_{N}=\vec{\nabla}\wedge \vec{J}_{N}\;.  \label{EBwe}
\end{equation}
The positron force is $-\vec{F}_{e}$. In a similar fashion, the neutrinos
suffer a weak force 
\begin{equation}
\vec{F}_{\nu }=\sqrt{2}G_{\mathrm{F}}c_{V}^{\prime }\,(\vec{E}_{e}+\vec{v}%
_{\nu }\wedge \vec{B}_{e})\;,\;  \label{Fn}
\end{equation}
($-\vec{F}_{\nu }$ for anti-neutrinos) where 
\begin{equation}
\vec{E}_{e}=-\vec{\nabla}J_{e}^{0}-\frac{\partial \vec{J}_{e}}{\partial t}%
\;,\qquad \vec{B}_{e}=\vec{\nabla}\wedge \vec{J}_{e}\;.  \label{EBwn}
\end{equation}
Above, the velocity and linear momentum are related to each other by $\vec{v}%
=\vec{p}/\sqrt{\vec{p}^{2}+m^{2}}$ with a zero mass in the neutrino case.

These weak forces differ from the ones employed before \cite
{bing94,silv98,lami99}. In particular, the ponderomotive forces considered
in \cite{bing94,silv98} only contain the terms proportional to the
gradients, $\vec{\nabla}J_{N}^{0}$ and $\vec{\nabla}J_{e}^{0}$, of the
neutrino and electron densities but not the terms that go with the vector
currents $\vec{J}_{N}$ and $\vec{J}_{e}$. In the case of the longitudinal
waves, the weak-magnetic forces are suppressed by an additional power of $G_{%
\mathrm{F}}^{2}$ but the time derivatives of $\vec{J}_{N}$ and $\vec{J}_{e}$
still contribute at the same level as the density gradients. This fact will
manifest in the dispersion relation itself.

The proper modes of a system are usually investigated by expanding the
distribution function around a static and uniform function $f^{0}$ \emph{i.e.%
}, 
\begin{equation}
f(x,\vec{p})=f^{0}(\vec{p})+\delta f(x,\vec{p})\;.  \label{df}
\end{equation}
The Vlasov equations (\ref{Dfen}) are then approximated to the linearized
form 
\begin{equation}
\left( \frac{\partial }{\partial t}+\vec{v}\cdot \!\frac{\partial }{\partial 
\vec{x}}\right) \delta f+\vec{F}\!\cdot \!\frac{\partial f^{0}}{\partial 
\vec{p}}=0\ ,
\end{equation}
which become \cite{lifs97} 
\begin{equation}
-i(\omega -\vec{k}\!\cdot \!\vec{v})\delta f(k,\vec{p})+\vec{F}(k,\vec{p}%
)\!\cdot \!\frac{\partial f^{0}}{\partial \vec{p}}=0\;
\end{equation}
after Fourier analysis, where $k^{\mu }=(\omega ,\vec{k})$ denote the
frequency and wave vector components. It is convenient to write this in a
relativistic covariant notation. In terms of 
\begin{equation}
F_{4}^{\mu }=p^{0}\frac{dp^{\mu }}{dt}\;,
\end{equation}
that transforms as a 4-vector, the equation above reads as 
\begin{equation}
-ik\!\cdot \!p\,\,\delta f(k,p)+F_{4}^{\mu }(k,p)\frac{\partial f^{0}}{%
\partial p^{\mu }}=0\;.
\end{equation}
It does not matter whether $f^{0}$ depends on $p^{0}$ or it is evaluated
on-shell ($p^{2}=m^{2}$) because $F_{4}^{\mu }\,\partial /\partial p^{\mu }$
is a total derivative when applied on functions that do not depend on the
space coordinates.

Under an external perturbation the current densities 
\begin{equation}
j^{\mu }(x)=\int \frac{d^{3}p}{(2\pi )^{3}E}\,f(x,p)\,p^{\mu }\   \label{jmu}
\end{equation}
($E$ is the kinetic energy) suffer fluctuations equal to 
\begin{equation}
j^{\mu }(k)=-i\int \frac{d^{3}p}{(2\pi )^{3}E}\frac{\partial f^{0}}{\partial
p^{\nu }}\frac{F_{4}^{\nu }(k,p)\,p^{\mu }}{k\!\cdot \!p}\ .  \label{djmu}
\end{equation}
To compare with the field theory results it is convenient to integrate by
parts obtaining 
\begin{equation}
j^{\mu }(k)=i\int \frac{d^{3}p}{(2\pi )^{3}E}f^{0}\frac{\partial }{\partial
p^{\nu }}\left( \frac{F_{4}^{\nu }\,p^{\mu }}{k\!\cdot \!p}\right) \ .
\label{jmuk}
\end{equation}
It must be kept in mind that here $\partial F_{4}^{\nu }/\partial p^{\nu }$
identifies with $E\,\partial \vec{F}/\partial \vec{p}$ after setting the
on-shell condition $p^{0}=E$. In the cases of interest to us the particles
interact with a vector potential, $V_{\mu }$, and the 4-force is 
\begin{equation}
F_{4}^{\mu }=(\partial ^{\mu }V^{\nu }-\partial ^{\nu }V^{\mu })p_{\nu }\;.\;
\end{equation}
Eqs.~(\ref{djmu}, \ref{jmuk}) yield a linear relation, $j^{\mu }=-\pi ^{\mu
\nu }V_{\nu }$, with susceptibility given by 
\begin{equation}
\pi ^{\mu \nu }[f^{0}]=\int \frac{d^{3}p}{(2\pi )^{3}E}\frac{\partial f^{0}}{%
\partial p^{\alpha }}\frac{k^{\alpha }p^{\nu }-k\!\cdot \!p\,g^{\alpha \nu }%
}{k\!\cdot \!p}p^{\mu }\;  \label{picl1}
\end{equation}
or, after integrating by parts, 
\begin{equation}
\pi ^{\mu \nu }[f^{0}]=-\int \frac{d^{3}p}{(2\pi )^{3}}\frac{f^{0}}{E}\frac{%
(k^{\mu }p^{\nu }+p^{\mu }k^{\nu }-k\!\cdot \!p\,g^{\mu \nu })k\!\cdot
\!p-k^{2}p^{\mu }p^{\nu }}{(k\!\cdot \!p)^{2}}\;.  \label{picl2}
\end{equation}

For instance, to obtain the electromagnetic polarization tensor it suffices
to make $J_{\mathrm{EM}}^{\mu }=q\,j^{\mu }$ and $V_{\nu }=qA_{\nu }$ for
each charged particle (charge $q)$ in the relation $J_{\mathrm{EM}}^{\mu
}=-\pi _{\mathrm{EM}}^{\mu \nu }A_{\nu }$. The result is a sum over all
particles and spin states of the terms $q^{2}\pi ^{\mu \nu }$. For the
electron plasma in particular (2 spin states) 
\begin{equation}
\pi _{\mathrm{EM}}^{\mu \nu }=2e^{2}\pi ^{\mu \nu }[f_{e}+f_{\bar{e}}]\;.
\end{equation}
It can be extended to the neutrino-plasma system simply by taking the vector
potentials indicated by Eqs.~(\ref{Le}, \ref{Ln}) namely, $-eA^{\mu }+\sqrt{2%
}G_{\mathrm{F}}c_{V}^{\prime }\,J_{N}^{\mu }$ for electrons and $\sqrt{2}G_{%
\mathrm{F}}c_{V}^{\prime }\,J_{e}^{\mu }$ for neutrinos. In this way, one
obtains the same relations as (\ref{JEM}, \ref{JN}) but with classic theory
susceptibilities given by 
\begin{equation}
\pi _{N}^{\mu \nu }=\pi ^{\mu \nu }[f_{\nu }+f_{\bar{\nu}}]\;  \label{pincl}
\end{equation}
and the tensor $\pi _{\mathrm{EM}}^{\mu \nu }$ above. The differences
between this and the field theory results (\ref{piem}, \ref{pin}) only
appear, vacuum corrections apart, in the particle propagators and
parity-violating terms. However, in the limit of frequencies and wavenumbers
much lower than the electron and neutrino energies, field theory delivers
the same results as classical kinetic theory. The dispersion relation of the
longitudinal waves will be analyzed next section .

\section{Waves and plasma instabilities.}

Bingham \emph{et~al.}~\cite{bing94,silv98} conceived a mechanism of energy
transfer from a neutrino beam to an electron plasma in which certain plasma
wave modes acquire large growth rates as a result of a neutrino resonant
effect. In order that those resonant waves be not electron Landau damped the
electron plasma has to be in a non-relativistic regime. Let $\omega _{pl}(%
\vec{k})$ designate the frequency of the longitudinal waves as a function of 
$\vec{k}$ in a plasma without neutrinos. In the case of a Maxwellian
distribution $\omega _{pl}$ is given by \cite{lifs97} 
\begin{equation}
\omega _{pl}^{2}(\vec{k})=\omega _{p}^{2}+3\frac{\omega _{p}^{2}}{k_{\mathrm{%
D}}^{2}}\vec{k}^{2}
\end{equation}
for wavenumbers much smaller than $k_{\mathrm{D}}$, the Debye wavenumber. $%
\omega _{p}$ is the plasma frequency and $\omega _{p}^{2}=4\pi \alpha
n_{e}/m_{e}$, $k_{\mathrm{D}}^{2}=4\pi \alpha n_{e}/T_{e}$ for an electron
density and temperature equal to $n_{e}$ and $T_{e}$, respectively. More
important to what follows is that for $k\ll k_{\mathrm{D}}$ (either
degenerate or non-degenerate gas) the frequency $\omega _{pl}$\ does not
vary much with $k$ and $\pi _{\mathrm{EM}}^{00}/\vec{k}^{2}$ is
approximately equal to $-\omega _{pl}^{2}/\omega ^{2}$, as long as $\omega /k
$ is much larger than $v_{eT}=\sqrt{T_{e}/m_{e}}$, the electron thermal
velocity. Using this relation in Eqs.~(\ref{drw}, \ref{piw00}, \ref{pin00})
one obtains the dispersion relation in the presence of neutrinos as 
\begin{equation}
\omega ^{2}-\omega _{pl}^{2}(\vec{k})=-\frac{G_{\mathrm{F}}^{2}c_{V}^{\prime
2}}{2\pi \alpha }\frac{\omega ^{2}}{\vec{k}^{2}}(\vec{k}^{2}-\omega
^{2})^{2}\pi _{N}^{00}\;,  \label{w2-w2}
\end{equation}
up to terms of $G_{\mathrm{F}}^{4}$ order (recall that $e^{2}=4\pi \alpha $%
), with 
\begin{equation}
\pi _{N}^{00}=-\int \frac{d^{3}p_{\nu }}{(2\pi )^{3}}\left( f_{\nu }+f_{\bar{%
\nu}}\right) E_{\nu }\frac{\vec{k}^{2}\!-(\vec{k}\!\cdot \!\vec{v}_{\nu
})^{2}}{(k\!\cdot \!p_{\nu })^{2}-(k^{2}/2)^{2}}\;.
\end{equation}

The expressions above put the real impact of the neutrino flux in
perspective. Keeping only the main factors, 
\begin{equation}
\omega ^{2}-\omega _{pl}^{2}\;\propto \;\omega _{pl}^{2}\,\frac{G_{\mathrm{F}%
}\,n_{\nu }}{E_{\nu }}\,G_{\mathrm{F}}\,\vec{k}^{2}\;
\end{equation}
clearly indicates that the neutrino contribution is severely suppressed by $%
G_{\mathrm{F}}^{2}$. The only potential exception is a strong neutrino
resonance effect. The poles in the neutrino propagators represent kinematic
conditions for a massless neutrino with momentum $p$ to emit or absorb a
plasmon with momentum $k$: $(p\pm k)^{2}=0=p^{2}$. A necessary condition for
such a \v{C}erenkov process is that $k$ be a space-like vector \emph{i.e.}, $%
\omega <|\vec{k}|$. That is quite possible for modes that are not Landau
damped, $|\vec{k}|<k_{\mathrm{D}}$, because the Debye wavenumber $k_{\mathrm{%
D}}$ of a non-relativistic plasma is much larger than the plasma frequency $%
\omega _{p}$.

The frequency and wavenumbers of interest are much smaller than the electron
and neutrino single particle energies. Therefore, it makes sense to neglect $%
k^{2}$ in front of $k\!\cdot \!p$ in the neutrino propagators so that

\begin{equation}
\pi _{N}^{00}=-\int \frac{d^{3}p_{\nu }}{(2\pi )^{3}}\frac{f_{\nu }+f_{\bar{%
\nu}}}{E_{\nu }}\frac{\vec{k}^{2}\!-(\vec{k}\!\cdot \!\vec{v}_{\nu })^{2}}{%
(\omega -\vec{k}\!\cdot \!\vec{v}_{\nu })^{2}}\;.  \label{pin00cl1}
\end{equation}
This is nothing but the classic theory result contained in the Eqs.~(\ref
{picl2}, \ref{pincl}), which gives also, taking Eq.~(\ref{picl1}) in
account, 
\begin{equation}
\pi _{N}^{00}=\int \frac{d^{3}p_{\nu }}{(2\pi )^{3}}\frac{1}{\omega -\vec{k}%
\!\cdot \!\vec{v}_{\nu }}\left( \frac{\partial f_{\nu }}{\partial \vec{p}}+%
\frac{\partial f_{\bar{\nu}}}{\partial \vec{p}}\right) \!\cdot \!\vec{k}\;.
\label{pin00cl2}
\end{equation}
Together with Eq.~(\ref{w2-w2}) it constitutes the dispersion relation
predicted with classical kinetic theory. Our result differs from others \cite
{silv98, lami99} simply because the forces assumed there to account for the
weak interactions are different from the Lorentz kind of force we derived in
section 3. Bingham \emph{et~al.}~in particular \cite{bing94,silv98}, only
considered the terms proportional to the electron and neutrino density
gradients and so obtained a factor of $(1-\omega ^{2}\!/\vec{k}^{2})^{2}$\
less in the dispersion relation \cite{silv98}. This factor comes from the
time derivatives of the currents $\vec{J}_{\nu }$ and $\vec{J}_{e}$ in the
forces (\ref{Fe}) and (\ref{Fn}): $\vec{J}=\omega \vec{k}J^{0}\!/\vec{k}^{2}$
for \emph{longitudinal} waves and the same type of relation holds for the
polarization tensor, Eq.~(\ref{pii0}), as a result of plasma isotropy and
current conservation. Altogether, it makes a factor of $1-\omega ^{2}\!/\vec{%
k}^{2}$, one for neutrinos and one for electrons, as can be learned from
Eqs.~(\ref{JEM}, \ref{JN}).

The kinetic instability is analogous to the electron Landau damping \cite
{lifs97} and takes place if the neutrino spectrum crosses from one side to
the other of the resonance ($\vec{k}\!\cdot \!\vec{v}_{\nu }=\omega $).
Then, the integral in Eq.~(\ref{pin00cl2}) separates into a principal part
and an imaginary quantity that is evaluated with the so-called Landau
prescription \cite{land87} in this case, with $\omega -\vec{k}\!\cdot \!\vec{%
v}_{\nu }+i0^{+}$ in the denominator. The neutrino contribution to the
damping rate comes then as 
\begin{equation}
\gamma _{_{\mathrm{W}}}=-\frac{\,G_{\mathrm{F}}^{2}\,c_{V}^{\prime \,2}}{%
4\alpha }\,\frac{\omega }{\vec{k}^{2}}\,(\vec{k}^{2}-\omega ^{2})^{2}\!\int
\!\frac{d^{3}p_{\nu }}{(2\pi )^{3}}\,\,\delta (\omega -\vec{k}\!\cdot \!\vec{%
v}_{\nu })\left( \frac{\partial f_{\nu }}{\partial \vec{p}}+\frac{\partial
f_{\bar{\nu}}}{\partial \vec{p}}\right) \!\cdot \!\vec{k}\ .  \label{gweak}
\end{equation}
Hardy and Melrose \cite{hard97} obtained this result starting from the decay
rate of single neutrinos into longitudinal photons. However, by the very
nature of such calculation the full neutrino contribution to the dispersion
relation was not derived, which might be important to investigate possible
reactive instabilities. 
Before going to that we just note that contrary to
the interpretation in \cite{hard97}, the factor of $(1-\omega ^{2}\!/\vec{k}%
^{2})^{2}$\ is not due to the chiral nature of weak interactions but rather
to their current-current structure, plasma isotropy
and a particularity of longitudinal waves.

$\gamma _{_{\mathrm{W}}}$ goes as $G_{\mathrm{F}}^{2}$ and is exceedingly
small when compared with the plasma collisional damping for instance. If
however all neutrinos were on the top of the resonance they could generate
an hydrodynamic instability. The claim was \cite{bing94,silv98} that this
occurs in the conditions of Supernova neutrino emission causing much larger
growth rates, proportional to $G_{\mathrm{F}}^{2/3}$ rather than $G_{\mathrm{%
F}}^{2}$, or to $G_{\mathrm{F}}$ if electron-ion collisions are taken into
consideration. The question we raise is, in the end one has to check whether
or not the entire neutrino flux lies in the resonance \emph{i.e.}, whether $%
|\omega -\vec{k}\!\cdot \!\vec{v}_{\nu }|$ is confined to the calculated
resonance width $|\gamma |$. If that is not so one falls in the Landau
damping case and result (\ref{gweak}).

The hydrodynamic limit is obtained by assuming that $\omega -\vec{k}\cdot 
\vec{v}_{\nu }$ is approximately constant over the neutrino spectrum in Eq.~(%
\ref{pin00cl1}) and then solving the dispersion relation for a complex $%
\omega $. A solution with positive imaginary part represents a reactive
instability. From Eqs.~(\ref{w2-w2}, \ref{pin00cl1}) one writes 
\begin{equation}
\omega ^{2}-\omega _{pl}^{2}(\vec{k})\simeq \frac{G_{\mathrm{F}}^{2}}{2\pi
\alpha }\left\langle \frac{n_{\nu }}{E_{\nu }}\right\rangle \frac{\omega ^{2}%
}{\vec{k}^{2}}(\vec{k}^{2}-\omega ^{2})^{2}\frac{\vec{k}^{2}\!-(\vec{k}%
\!\cdot \!\vec{v}_{\nu })^{2}}{(\omega -\vec{k}\!\cdot \!\vec{v}_{\nu })^{2}}%
\;,
\end{equation}
where $n_{\nu }$ means the joint $\nu _{e}$ and $\bar{\nu}_{e}$ flux and we
have made $c_{V}^{\prime }=1$. The largest growth rates ($\mathrm{Im}%
\{\omega \}=-\gamma >0$) are obtained for resonant modes ($\vec{k}\!\cdot \!%
\vec{v}_{\nu }\approx \omega _{pl}$) with magnitudes around 
\begin{equation}
\Gamma =\omega _{pl}\left\{ \frac{G_{\mathrm{F}}\,n_{\nu }}{E_{\nu }}\frac{%
G_{\mathrm{F}}\,n_{e}}{m_{e}}\right\} ^{1/3}\frac{\vec{k}^{2}-\omega
_{pl}^{2}}{\omega _{pl}^{4/3}\,\vec{k}^{2/3}}\;.
\end{equation}
But this assumes that $\vec{k}\cdot \vec{v}_{\nu }$ covers an interval of
width $\,\Delta \vec{k}\!\cdot \!\vec{v}_{\nu }$ not greater than $\Gamma $.
The neutrino velocities only spread in direction. Far away from the
neutrinosphere of radius $R_{\nu }$ they essentially move in the radial
direction yet, the velocity cone at a radius $r$ has a finite aperture, $%
\theta _{\nu }\approx 2R_{\nu }/r$. Hence, 
\begin{equation}
\Delta \vec{k}\!\cdot \!\vec{v}_{\nu }\approx \theta _{\nu }\,k\,\sin \theta
\,\approx \theta _{\nu }\,\omega _{pl}\left| \tan \theta \right| \,
\end{equation}
for the resonant modes, where $\theta $ is the angle $\vec{k}$ makes with
the radial direction. $\Gamma $ increases with $\tan ^{4/3}\theta $ so the
best chance of satisfying the requirement $\Gamma >\Delta \vec{k}\!\cdot \!%
\vec{v}_{\nu }$ is to have an angle $\theta $ as close as possible to $\pi /2
$. It must however not get any closer than $\pi /2\pm \theta _{\nu }$
otherwise $\Delta \vec{k}\!\cdot \!\vec{v}_{\nu }$\ becomes larger than $%
\omega _{pl}$ and neutrino Landau damping cannot be avoided. That puts an
upper limit on the ratio $\Gamma /\Delta \vec{k}\!\cdot \!\vec{v}_{\nu }$
which implies the necessary condition 
\begin{equation}
\hat{\gamma}_{1}=\left\{ \frac{G_{\mathrm{F}}\,n_{\nu }}{E_{\nu }}\frac{G_{%
\mathrm{F}}\,n_{e}}{m_{e}}\right\} ^{1/3}\theta _{\nu }^{-4/3}>1\;.
\end{equation}
Another upper limit and necessary condition comes from that $k$ must be
smaller than the Debye wavenumber. The necessary condition is 
\begin{equation}
\hat{\gamma}_{2}=\left\{ \frac{G_{\mathrm{F}}\,n_{\nu }}{E_{\nu }}\frac{G_{%
\mathrm{F}}\,n_{e}}{m_{e}}\right\} ^{1/3}\frac{k_{\mathrm{D}}}{\theta _{\nu }%
}>1\;.
\end{equation}

Knowing how small the energies $G_{\mathrm{F}}n_{\nu }$ and $G_{\mathrm{F}%
}n_{e}$ are those conditions look quite disfavored by data. Take a neutrino
luminosity \cite{mayl87} $L_{\nu }=10^{53}\,\mathrm{ergs/s}$, $E_{\nu }=10\,%
\mathrm{MeV}$, neutrinosphere radius $R_{\nu }=30\,\mathrm{km}$ and a
generous electron density $n_{e}=10^{30}\,\mathrm{cm}^{-3}$, barely
compatible with the non-relativistic regime. Recalling that $\theta _{\nu
}=2R_{\nu }/r$ and $n_{\nu }=L_{\nu }/4\pi r^{2}$ one gets 
\begin{equation}
\hat{\gamma}_{1}=1.26\left( \frac{r}{10^{14}\,\mathrm{km}}\right) ^{2/3}\;.
\end{equation}
It means that for an electron density as high as $10^{30}\,\mathrm{cm}^{-3}$
a strong $\nu $ resonance effect could only take place at a radius of $%
10^{14}\,\mathrm{km}\ $or larger! That is clearly absurd.\ The conclusion we
draw is the neutrinos are never collimated enough to cause an hydrodynamic
instability. The resonance width is far too small for that. It slices the
neutrino velocity cone in two pieces, one above and one below the resonance $%
\vec{k}\!\cdot \!\vec{v}_{\nu }=\omega $. The neutrinos only induce Landau
damping with a rate given by the expression (\ref{gweak}). This was
calculated by Hardy and Melrose \cite{hard97} for an electron density of $%
10^{30}\,\mathrm{cm}^{-3}$ and they found that the growth rates are too
small for the time duration of neutrino emission in Supernovae. When one
tries to push the electron density to increase the growth rates one
approaches the relativistic regime but then electron Landau damping takes
over. In addition, the phase space of plasma waves with phase velocity less
than one ($\omega <|\vec{k}|$, necessary for \v{C}erenkov neutrino emission)
becomes vanishing small thus suppressing the relative factor $(1-\omega
^{2}\!/\vec{k}^{2})^{2}$ in the dispersion relation. On the other hand, the
collisional damping in a non-relativistic plasma is many orders of magnitude
larger than the\ neutrino Landau damping. To conclude, it looks that the
neutrinos are not capable of transferring any significant energy to the
medium through plasma wave instabilities.

\section{Conclusions and discussion}

We applied the techniques of finite temperature field theory to study the
longitudinal modes of the electromagnetic waves in a plasma crossed by an
intense neutrino flux. It is shown that for an isotropic plasma the electron
axial-vector couplings and polarization effects are suppressed by $G_{%
\mathrm{F}}^{4}$ in the dispersion relation. Only the electron vector
couplings (weak and electromagnetic) contribute at $G_{\mathrm{F}}^{2}$
order. In addition, in the limit of frequencies and wavenumbers much smaller
than the individual electron and neutrino energies, the susceptibility
tensors derived from field theory are well approximated by the results
obtained with classic kinetic theory provided that the electroweak
interactions are described with the appropriate forces. As a result of the
vectorial nature of the interactions, the weak forces are of the same type
as the Lorentz electromagnetic force. They diverge however from the forces
employed by other authors \cite{bing94, hard96, silv98, lami99} in
particular the ponderomotive force of Bingham \emph{et~al.}~\cite{bing94,
silv98}. That explains the difference between the dispersion relation
derived by us and the ones obtained in \cite{silv98} and \cite{lami99} using
classic kinetic theory.

In the early papers of Bingham \emph{et~al.}~\cite{bing94} the neutrinos
were treated with a wave function obeying a sort of Klein-Gordon equation
with an external potential accounting for the interaction with the medium.
The collective effects were attributed there to a puzzling phase coherence
between the neutrino wave functions. There is nothing wrong with that: it
simply means that under an external plasma wave of wavenumber $k^{\mu }$ the
neutrino wave function fluctuations share a phase factor, common to all
neutrinos ($e^{-ik\cdot x}$), on top of the arbitrary initial phases of each
neutrino wave function. The problem was rather that in this Klein-Gordon
sort of equation (which could in principle be derived by squaring the Dirac
equation) all the terms that might depend on the external potential
derivatives or electron 3-vector current were completely discarded. That is
fine to study problems like neutrino oscillations but not for neutrino
effects on plasma waves because their very nature concerns the variations of
the medium densities on the time and length scales of the wave period and
wavelength.

The other problem concerns the claimed \cite{bing94, silv98} non-linear
effects and plasma instabilities induced by resonant neutrinos ($\vec{k}%
\!\cdot \!\vec{v}_{\nu }=\omega $) on a non-relativistic plasma. We compared
the calculated damping (growth) rate of the plasma waves with the phase
space occupied by the neutrino flux and concluded that the angular
dispersion of the neutrino velocity due to the finite size of the
neutrinosphere exceeds by far the assumed resonance width, at least a factor
of $10^{8}$\ even in the case of the most optimistic growth rate, no matter
how large is the distance to the neutrinosphere. This is of course due to
the severe weakness of the neutrino interactions. As a consequence no
hydrodynamic (reactive) instabilities can be induced by neutrinos. What is
left is just neutrino Landau damping, a linear effect proportional to $G_{%
\mathrm{F}}^{2}$. The corresponding growth rates are however too small for
the time duration of neutrino emission in Supernovae \cite{hard97}. In
addition they are overcame by either collisional damping in a
non-relativistic plasma or by electron Landau damping in a relativistic one.
In conclusion, wave instabilities induced by neutrinos do not seem to be a
viable mechanism of substantial energy transfer from neutrinos to a
Supernova plasma.

\section*{Acknowledgments}

I thank Tito Mendon\c{c}a, Lu\'{\i }s Silva and Ana Mour\~{a}o for valuable
discussions. This work was supported in part by FCT under the grant
PESO/P/PRO/1250/98.


\end{document}